# PROMOTING EFFECTIVE APPLICATION AND MANAGEMENT OF ICT TO ENHANCE PERFORMANCE IN SECONDARY SCHOOLS

Palagolla W. W. N. C. K.
*Lecturer, Department of Accountancy and Finance, Rajarata University, Sri Lanka*

Wickramarachchi A. P. R.
*Senior Lecturer, Department of Industrial Management, University of Kelaniya, Sri Lanka*

**ABSTRACT**

Information and Communication Technology (ICT) is widely used to enhance teaching and learning in modern education. However, it is still difficult in some countries to exploit the full potential of ICT in this regard due to various constraints. The paper explores existing barriers for the effective use of ICT to enhance performance of secondary schools. The study was focused on secondary schools in the North Central Province (NCP) of Sri Lanka. Results show a very low use of ICT among schools in the territory due to poor ICT infrastructure, leadership support, school planning, and ICT competency. However, teachers' fairly positive attitudes towards ICT indicated a positive aspect on future developments. Individuals' demographic characteristics show significant differences in the use of ICT in schools. Moreover, positive relationships were found between ICT usage and performance as perceived by teachers.

## 1. INTRODUCTION

Use of ICT is growing rapidly over diverse organizations due to its enormous potential for enhancing organizational performance. Although the power of ICT has penetrated into many sectors, the secondary education especially in the rural Sri Lanka seems to lag behind compared to other countries with similar economic and social background. In response, Sri Lanka like many other countries is making efforts to promote ICT in the secondary schools as a long term strategy to minimize digital divide.

In recent years, many schools have moved rapidly towards integrating computer based technologies in order to enhance teaching and learning experience. Teachers in UK regard ICT as an essential skill in education and report increased use of computers for planning and preparing lessons, administration, assessment and tracking students' progress (Irvine and Williams, 2002). Neo (2005) pointed out students' positive attitudes towards multimedia based learning environment increased interests in subject matter. Chen et al. (2009) noted that 'online schools' in China mainly deliver their courses through a satellite - based TV net or the Internet. M-learning has also emerged as an educational application and a survey in UK revealed students with the ages between 16 to 24 years are switched to learning by mobile phones and PDAs (Muyinda, 2007). Learning platforms and e-portfolios enable a range of tools for communication, collaboration, content management, curriculum planning etc. in schools.

Past research show effective use of ICT can enhance, extend, and enable teacher - student interactions. As noted by Tinio (2003), appropriate use of ICT raises the quality of education. Uhomoibhi (2006) indicated E-learning as a powerful tool in shaping teaching – learning process since it provides new ways for knowledge transfer. Also technology could result in increased self-confidence and eagerness to learn (Kimble, 1999). Moreover, Nordic (2006) pointed out ICT can positively impact on learning through engagement, differentiation, creativity and less waste of time. Newhouse et al. (2002) mentioned that effective application and management of ICT in classrooms lead active learning, productivity, motivation, higher level thinking, and independence that enable teachers to adopt a balanced pedagogical approach between teacher-centered instruction and learner-centered collaborative environment. Although, there is an increasing importance of the use of ICT, many basic questions about the use of ICT in schools around the world remain largely unanswered (Trucano, 2009). Furthermore, despite strong body of research showing great impact of ICT in schools, difficult and ineffective ICT integration is still common (Fluck, 2003; UNESCO, 2009). This paper identifies existing barriers towards effective ICT usage which in turn enhances performance of the secondary schools in the NCP of Sri Lanka.

Successful utilization largely depends on users especially teachers awareness, competency, attitudes and perceptions on ICT use as well as leadership support, culture and background, school planning, and availability of infrastructure (Ali and Proctor, 2005; Newhouse, 2002; Schiller, 2003; Yuen et al., 2003). Hence, a better understanding of prerequisites for the use of ICT in schools would be an important input in preparing schools for the 21$^{st}$ century challenges. Therefore, the findings would be extremely significant for stakeholders of the secondary education in order to address existing problems effectively.

## 2. METHODOLOGY

The paper investigates individual and organizational factors affecting use of ICT in schools which were identified through previous similar studies (Becta, 2007; Granger et al., 2002; Mumtaz, 2000) and preliminary survey. It also analyses the impact of ICT on work performance. Teachers' ICT competency and attitudes towards ICT are two major individual factors considered in the study. The potential organizational factors are ICT infrastructure, school planning and leadership support. Further, the study investigates the impact of individuals' demographic characteristics on the use of ICT in schools. The selected sample consists of 145 teachers drawn randomly from 30 ICT facilitated secondary schools in the NCP of Sri Lanka. A structured questionnaire rated on a five point Likert scale used to collect empirical data.

## 3. RESULTS AND DISCUSSIONS

Through mean score (M) analysis, it was identified that usage of ICT for various activities in schools is relatively very low (see Table 1). Frequencies also show that a very low usage of ICT among majority of teachers while approximately 16 percent reported a higher use of ICT. On average, 10 – 14 percent of staff indicated a medium level ICT use in school activities.

Table 1. Teachers' ICT usage in school

| ICT Usage | M | Staff Percentages | | |
|---|---|---|---|---|
| | | High | Medium | Low |
| Teaching & related activities | 2.1 | 15.9 | 11.1 | 73.1 |
| Assessment (Grading) | 2.0 | 15.9 | 13.1 | 71.0 |
| Professional development | 2.1 | 16.6 | 13.8 | 69.7 |
| Communication & collaboration | 1.96 | 15.2 | 9.6 | 75.2 |
| Decision making | 1.9 | 15.9 | 10.3 | 73.8 |
| Data/information storage (backup) | 1.9 | 15.9 | 9.7 | 74.5 |

Scale: Low (M < 2.5), Moderate (M < 3.5), and High (M ≥ 3.5)

### 3.1 Barriers towards Effective ICT Usage

Although different actions have been taken to promote effective ICT usage among secondary schools of Sri Lanka, findings highlighted that still ICT is under-utilized in secondary schools of the territory. In order to identify reasons behind the low usage, impact of identified factors (see Table 2) that affect the ICT usage was analyzed. In addition, perceived relationship between these factors and ICT usage was derived. As shown by correlations analysis all factors show strong positive relationships with ICT usage. However, findings indicated that unsatisfactory levels of appropriate 'ICT infrastructure', provision of 'leadership support' and proper 'planning to use ICT'. These factors are acting as barriers to effective ICT utilization. Results also showed moderate 'ICT competency' and moderately high 'attitudes towards ICT'. A moderate level of 'ICT competency' may also indicate that the government's efforts in improving ICT competency levels among teachers is not very successful.

Table 2. Factors affecting effective use of ICT

| Variable (s) | R | M |
|---|---|---|
| ICT Infrastructure | 0.82** | 2.09 |
| Leadership Support | 0.80** | 2.04 |
| School Planning | 0.78** | 2.03 |
| ICT Competency | 0.75** | 2.79 |
| Attitudes towards ICT | 0.70** | 3.10 |

Scale: Low (M < 2.5), Moderate (M < 3.5), and High (M ≥ 3.5)
Correlation: Weak (r < 0.5), Moderate (r = 0.5), Strong (r > 0.5)
Correlations are significant at: *p< 0.05, **p< 0.01

## 3.2 Group Differences in ICT Usage

The paper also analyzes through the use of ICT for various school activities listed in the Table 1 against attributes of teachers (see Table 3). It could be shown that teachers younger than 35 years are more inclined to use ICT than older ones. This point is also corroborated by service periods. Teachers with longer service period appear to use relatively low levels of ICT. This phenomenon could generally be observed in other sectors as well. One reason for this could be that ICT was introduced into schools curricula recently. Another reason may be that young people are generally more comfortable with technology. The analysis also show that male staff tends to use more ICT than female staff. This is in line with previous studies which also indicate existence of significant gender differences regarding attitudes toward ICT, extent of computer use, and self-perceived computer experience favoring males (Broos, 2005; Kwapong, 2009).

Table 3. Group differences in ICT usage

| Attributes of teachers | Teaching | Assessments | Decision Making | Communication & Collaboration | Professional Development | Data/info. Backup |
|---|---|---|---|---|---|---|
| Age: ≥ 35 (older staff) | 1.80 | 1.80 | 1.73 | 1.77 | 1.77 | 1.71 |
| <35 (younger staff) | 2.64 | 2.69 | 2.42 | 2.57 | 2.68 | 2.39 |
| Gender: Male | 2.22 | 2.22 | 2.10 | 2.17 | 2.24 | 2.06 |
| Female | 1.88 | 1.85 | 1.76 | 1.85 | 1.84 | 1.76 |
| Service: ≥ 15 | 1.73 | 1.69 | 1.69 | 1.72 | 1.70 | 1.70 |
| < 15 | 2.30 | 2.31 | 2.13 | 2.24 | 2.31 | 2.09 |
| Cycle of Teaching: | | | | | | |
| 6-11 Grade | 1.84 | 1.83 | 1.74 | 1.80 | 1.81 | 1.70 |
| 12-13 Grade | 2.52 | 2.52 | 2.39 | 2.51 | 2.57 | 2.40 |
| ICT Education: Yes | 2.41 | 2.41 | 2.23 | 2.38 | 2.45 | 2.25 |
| No | 1.74 | 1.72 | 1.68 | 1.69 | 1.69 | 1.63 |
| ICT Training: Yes | 2.07 | 2.08 | 1.89 | 2.03 | 2.09 | 1.90 |
| No | 2.05 | 2.02 | 1.99 | 2.02 | 2.02 | 1.94 |
| English Language Proficiency: High | 2.69 | 2.73 | 2.48 | 2.65 | 2.70 | 2.49 |
| Medium | 2.17 | 2.16 | 2.04 | 2.13 | 2.16 | 2.00 |
| Low | 2.15 | 1.54 | 1.54 | 1.54 | 1.56 | 1.51 |

Results also show that ICT utilization is high among staff teaching in grade 12 and grade 13. In general, teachers who serve upper grades possess higher academic qualifications than who serve for the lower grades.

They are likely to expose into more formal ICT education than others. This also validated by formal ICT education received by teachers. Accordingly, teachers with ICT education use more ICT for various school activities. However, it is worth noting that only a slight difference was observed between staff with and without ICT training. This may be due to inadequate and/or inappropriate training opportunities provided. Table 1 also highlighted the low 'ICT competency' levels among staff. Majority of respondents have commented that the existing training programs provide a limited support on how to integrate ICT into activities at work. It was also found a positive relationship between use of ICT and the level of English language proficiency of teachers.

## 3.3 Effective ICT Usage and Work Performance

ICT is universally accepted as a powerful enabler of quality, efficiency, and effectiveness of work which are commonly known as performance indicators. These dimensions were measured using number of variables, namely: work efficiency, saving of work resources, quantity of work, quality of work, achieving career goals, improving students' results, improving students' learning interest, and improving students' creativity. Results reveal that ICT usage is strongly and positively associated with each of the performance indicators (see Table 4). Accordingly performance of individual teachers and as well as school related activities could be improved significantly by better use of ICT. Results presented above indicate that low use of ICT among majority of teachers. However, results provided by table 4 indicate that they also recognize the importance of ICT in relation to improving performance. Thus, it could be assumed that school performance could be improved with appropriate facilities and support etc.

Table 4. Impact of ICT usage on work performance

| Performance Indicators | R |
|---|---|
| Work efficiency | 0.79** |
| Saving work resources | 0.78** |
| Quantity of work | 0.80** |
| Quality of work | 0.79** |
| Achieving career goals | 0.80** |
| Improving students' results | 0.76** |
| Improving students' learning interest | 0.75** |
| Improving students' creativity | 0.76** |

Correlation: Weak ($r < 0.5$), Moderate ($r = 0.5$), Strong ($r > 0.5$)
Correlations are significant at: *$p < 0.05$, **$p < 0.01$

## 4. CONCLUSION

The paper revealed that ICT usage in secondary schools of the NCP, Sri Lanka is still unsatisfactory owing to number of potential barriers such as insufficient ICT infrastructure, lack of leadership support and improper planning both at school level and higher levels. Another important finding is the limited success of training programmes in ICT. In addition, it was also found that teachers recognized that better utilization of ICT lead to improved performance.

Therefore, utilization of ICT and also overall performance of schools could be improved by proving adequate level of ICT infrastructure, organizing more effective training programmes in ICT, improving English language proficiency, and generally increasing the awareness among staff regarding usage of ICT tools in schools. Furthermore, school administrators providing a better leadership and support role may also lead to an enhanced use of ICT. Introduction of periodic reviews, benchmarking, and activities to motivate teachers and students to use ICT could also lead to achieving better results.

## 4.1 Limitations and Future Research

The findings could not be able to generalize to the entire country owing to uneven economic and social conditions among provinces. Further, the research can be improved by studying more aspects affecting effective use of ICT in schools. Longitudinal replications at national level will be useful to validate issues identified and formulate strategies to address them.